\documentclass{ws-procs9x6}
\usepackage{graphicx}

\newcommand{\intvecx}{\int d^3 x\,}

\newcommand{\veck}{{\bf k}}

\newcommand{\vecx}{{\bf x}}       
\newcommand{\vecy}{{\bf y}}
\newcommand{\vecz}{{\bf z}}

\newcommand{\dl}{\delta}

\newcommand{\kp}{\kappa}
\newcommand{\lm}{\lambda}

\newcommand{\ph}{\phi}

\newcommand{\om}{\omega}

\newcommand{\half}{\frac{1}{2}}

\newcommand{\Tr}{\mbox{Tr}\,}

\newcommand{\eela}[1]{\label{#1}\end{equation}}
\newcommand{\eeala}[1]{\label{#1}\end{eqnarray}}
\newcommand{\be}{\begin{equation}}
\newcommand{\ee}{\end{equation}}
\newcommand{\bea}{\begin{eqnarray}}
\newcommand{\eea}{\end{eqnarray}}

\begin{document}

\title{W-Particle Distribution in ElectroWeak Tachyonic
Pre-Heating\footnote{\uppercase{P}resented by \uppercase{J}.\ \uppercase{S}mit}}

\author{Jonivar Skullerud, Jan Smit 
and Anders Tranberg}

\address{Institute for Theoretical Physics, University of Amsterdam\\
Valckenierstraat 65, 1018 XE Amsterdam, the Netherlands}

\maketitle

\abstracts{Results are presented of a numerical study of the distribution of
W-bosons generated in a tachyonic electroweak pre-heating transition.}

\section{W-distribution in tachyonic electroweak baryogenesis}
The electroweak transition may have taken place shortly after inflation
ending at a low energy scale, and this possibility has been used to
suggest alternative scenarios for baryogenesis.
We have studied such a transition in the SU(2)-Higgs model 
with effective CP-violation \cite{ST}. The transition is assumed to
have taken place at zero temperature.
In the quenching approximation, it is induced simply by flipping the sign
of the quadratic term in the Higgs potential,
$\mu^2 \ph^{\dagger}\ph \to -\mu^2 \ph^{\dagger}\ph$. This causes the
Higgs field to go through a
spinodal instability in which its particle numbers initially grow
exponentially fast, leading to classical behavior 
\cite{Felder:2000hj,Smit:2001qa,Garcia-Bellido:2001cb,Copeland:2002ku,Garcia-Bellido:2002aj,Boea}. 
The gauge fields react strongly, and it is
interesting to see the emergence of effective W-particles, their
energy spectrum $\om_k$, and their distribution function $n_k$.

The numerical simulation is carried out in the temporal gauge $A_0=0$.
To define $\om_k$ and $n_k$, we transform to the Coulomb gauge
$\partial_j A_j = 0$ (which is a smooth gauge in which it makes sense
to Fourier transform to momentum space),
and maximally
coarse-grain the equal-time correlators of the gauge field
and its canonical conjugate ($E_j$) over the (periodic) volume
$L^3$,
\bea
\frac{1}{3L^3}\int d^3 z\, \langle A^p_i(\vecx + \vecz)
A^p_j(\vecy + \vecz)\rangle &=& C^{AA}_{ij}(\vecx-\vecy),
\nonumber\\
\frac{1}{3L^3}\int d^3 z\, \langle E^p_i(\vecx + \vecz)
E^p_j(\vecy + \vecz)\rangle &=& C^{EE}_{ij}(\vecx-\vecy),
\nonumber
\eea
where we also averaged over the three isospin directions.
After Fourier transformation, averaging over the transverse modes,
\bea
\half\dl_{ij}\intvecx e^{-i\veck\cdot\vecx}\, C^{AA}_{ij}(\vecx)
&=& C^{AA}_\veck,
\nonumber\\
\half\left(\dl_{ij} - \hat k_i\hat k_j\right)
\intvecx e^{-i\veck\cdot\vecx}\, C^{EE}_{ij}(\vecx) &=& C^{EE}_\veck,
\nonumber
\eea
and averaging over directions, $\veck \to k=|\veck|$,
the time-dependent particle spectrum and distribution function is obtained by
solving
\[
C^{AA}_k = n_{k}/\om_{k}
\;\;\;\;\; \;\;\;\;\;\;\;
C^{EE}_k = n_{k} \om_{k}
\]
for $n_k$ and $\om_k$.
Note that $n_k$ appears and not $n_k + 1/2$, since we are using the classical
approximation.
\begin{figure}[ht]
\includegraphics*[width=0.49\textwidth]{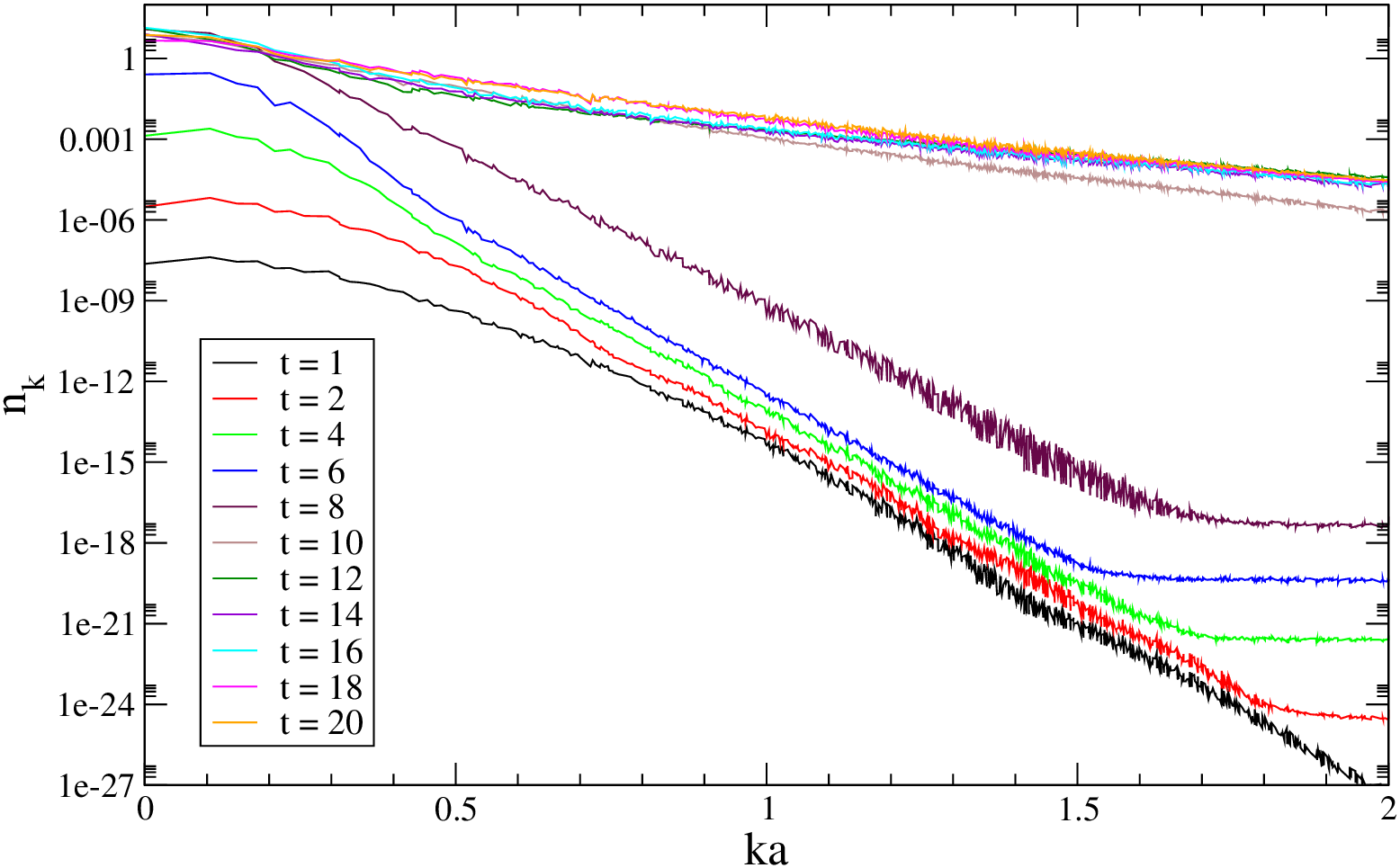}
\includegraphics*[width=0.49\textwidth]{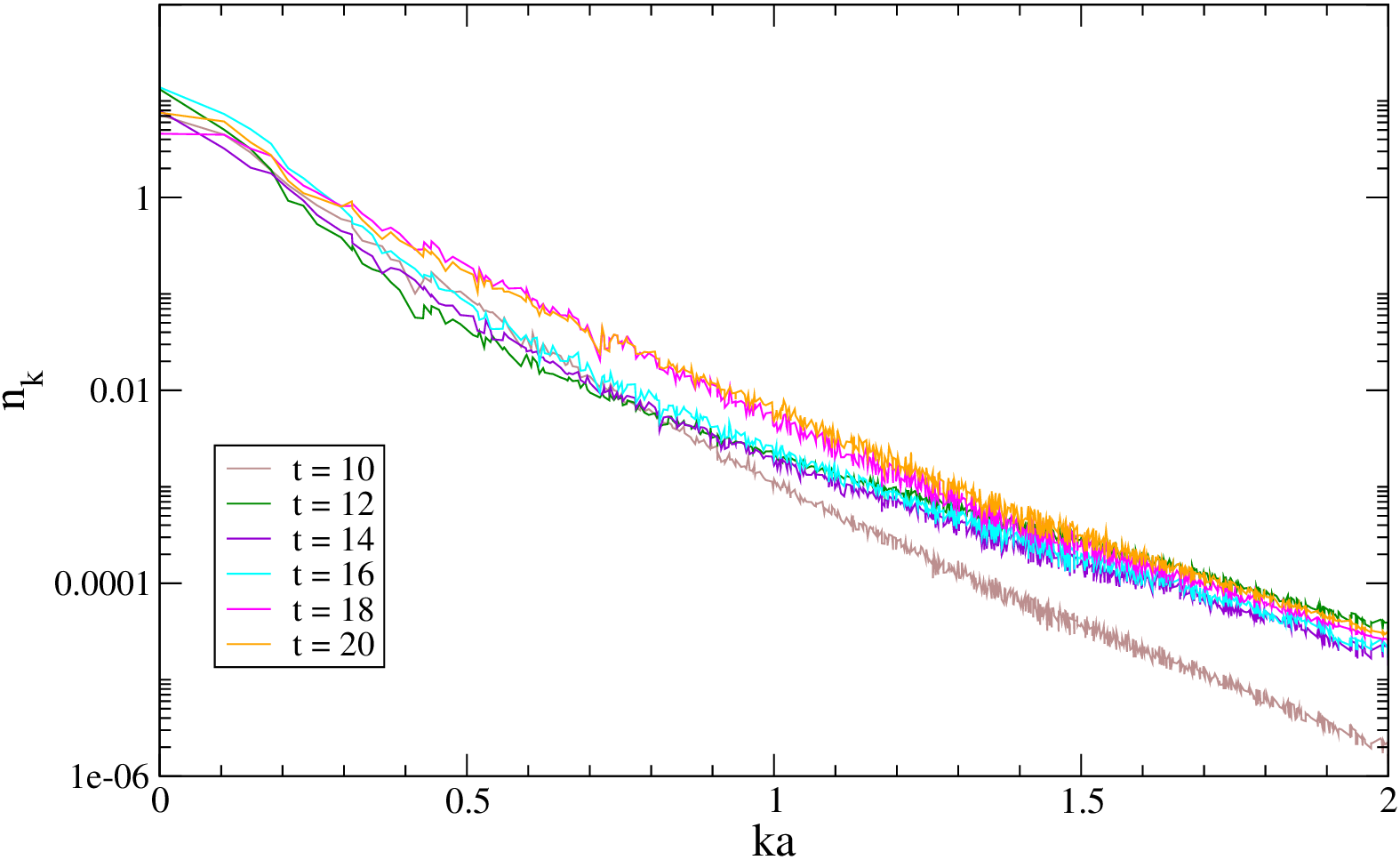}
\caption{
$n_k$ versus $ak$ at early times. Time is in units of $m_H^{-1}$
and the quench is at $t=0$.
}
\end{figure}

We present here results obtained from only {\em one}
gauge field configuration, i.e.\ the average over initial conditions
$\langle \cdots\rangle$ is not carried out. If classical behavior
applies, then one field configuration corresponds to one
realized universe. Of course, statistical observables such as $n_k$
and $\om_k$ will suffer from `cosmic variance', which may be severe in
a small volume.
The parameters of the simulation
on a lattice of $60^3$ sites are given by
\bea
\mbox{couplings:}
&&g^2 = 4/9,\;\;\;
1= 8\lm/g^2 = m_H^2 / m_W^2
\nonumber\\
\mbox{physical size}
&&L = 21\, m_H^{-1},\;\;\;
\mbox{lattice spacing}\;\;
a = 0.35\, m_H^{-1}
\nonumber
\eea

Figure 1 shows $n_k$ at early times after the quench at $t=0$.
We see an exponential growth of the low momentum modes until they saturate
around $t \approx 8\, m_H^{-1}$, after which the tail quickly acquires
a different slope on the log-plot.
\begin{figure}[ht]
\center{
\includegraphics*[width=0.70\textwidth]{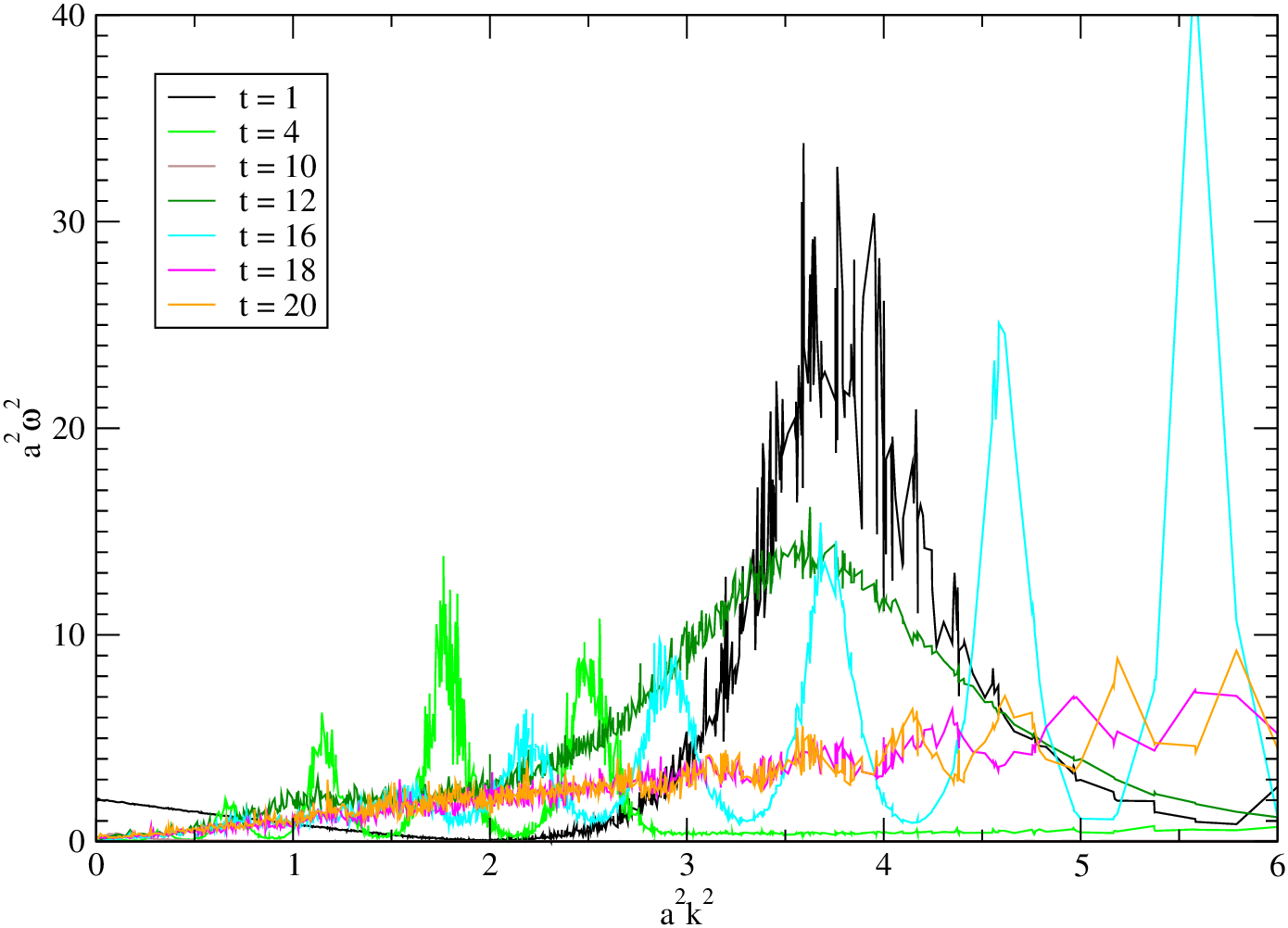}
}
\caption{
$a^2\om_k^2$ versus $a^2 k^2$ at early times.
The particle interpretation 
starts to make sense from $tm_H \approx 18$ onwards.
}
\end{figure}
\begin{figure}
\includegraphics*[width=0.49\textwidth]{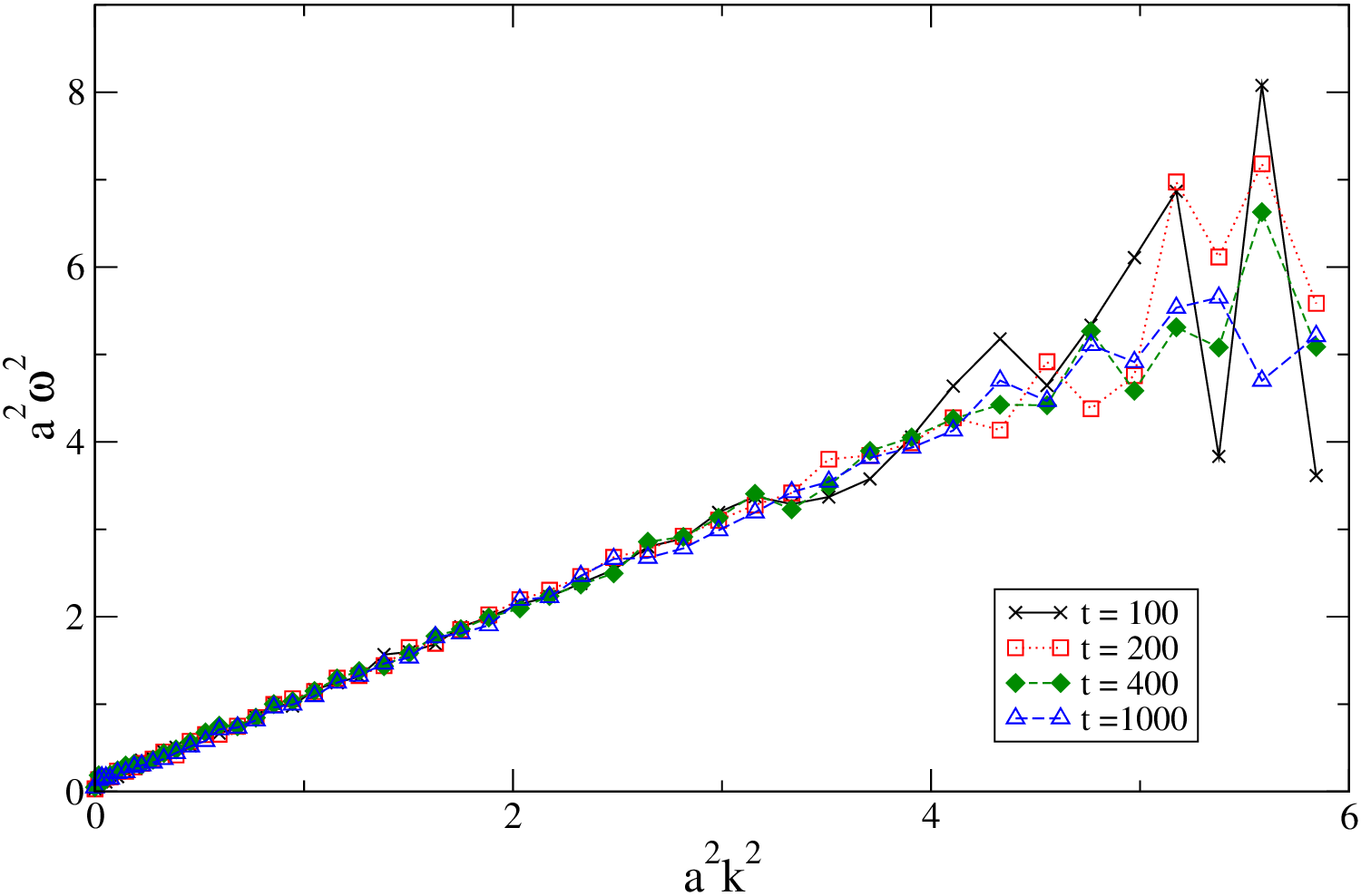}
\includegraphics*[width=0.49\textwidth]{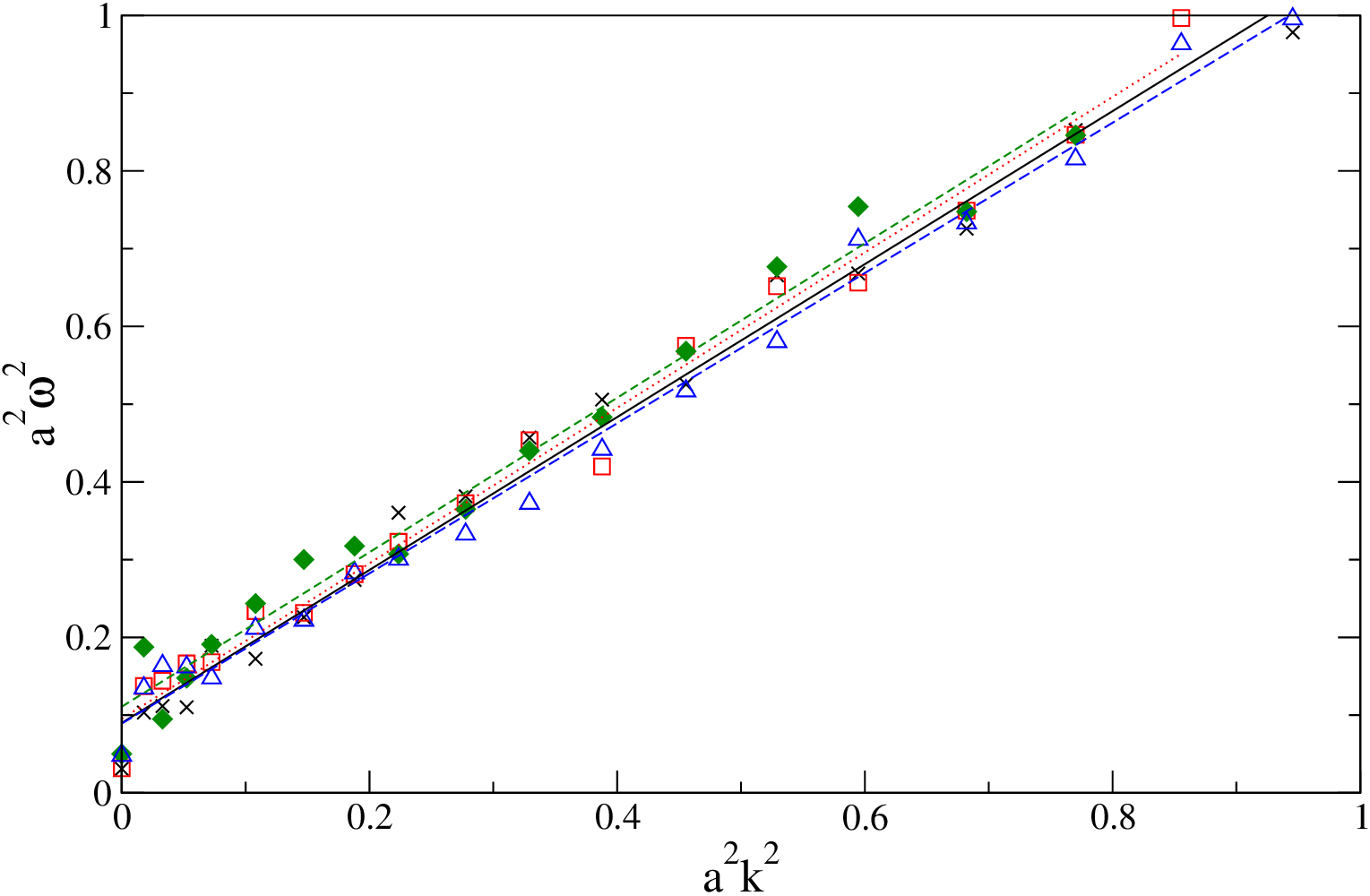}
\caption{
$a^2\om_k^2$ versus $a^2 k^2$ at later times.
The fitted slopes are 1 within a few percent.
The fitted
intercepts correspond to effective masses $m_{\rm eff}/m_H \approx 0.86$,
0.88, 0.95, 0.85 respectively for $tm_H = 100$, 200, 400, 1000.
}
\end{figure}
Figure 2 shows $\om_k$ for $t \leq 20\, m_H^{-1}$. It suffers greatly from
fluctuations (`cosmic variance' in our small `universe'), and we
averaged the data in bins of size $\Delta |\veck| \approx 0.14\, m_H$.  
Only from $t = 18\, m_H^{-1}$ onwards
a sensible $\om_k$ emerges, as can also be seen at later times in Figure 3.
The spectrum is close to the simple form $\om_k^2 = m_{\rm eff}^2 + k^2$
(on the lattice we use $k^2 \to a^{-2}\sum_j [2-2\cos(ak_j)]$),
with $m_{\rm eff} \approx m_W$, the zero-temperature input-value.
\begin{figure}
\center{
\includegraphics*[width=0.70\textwidth]{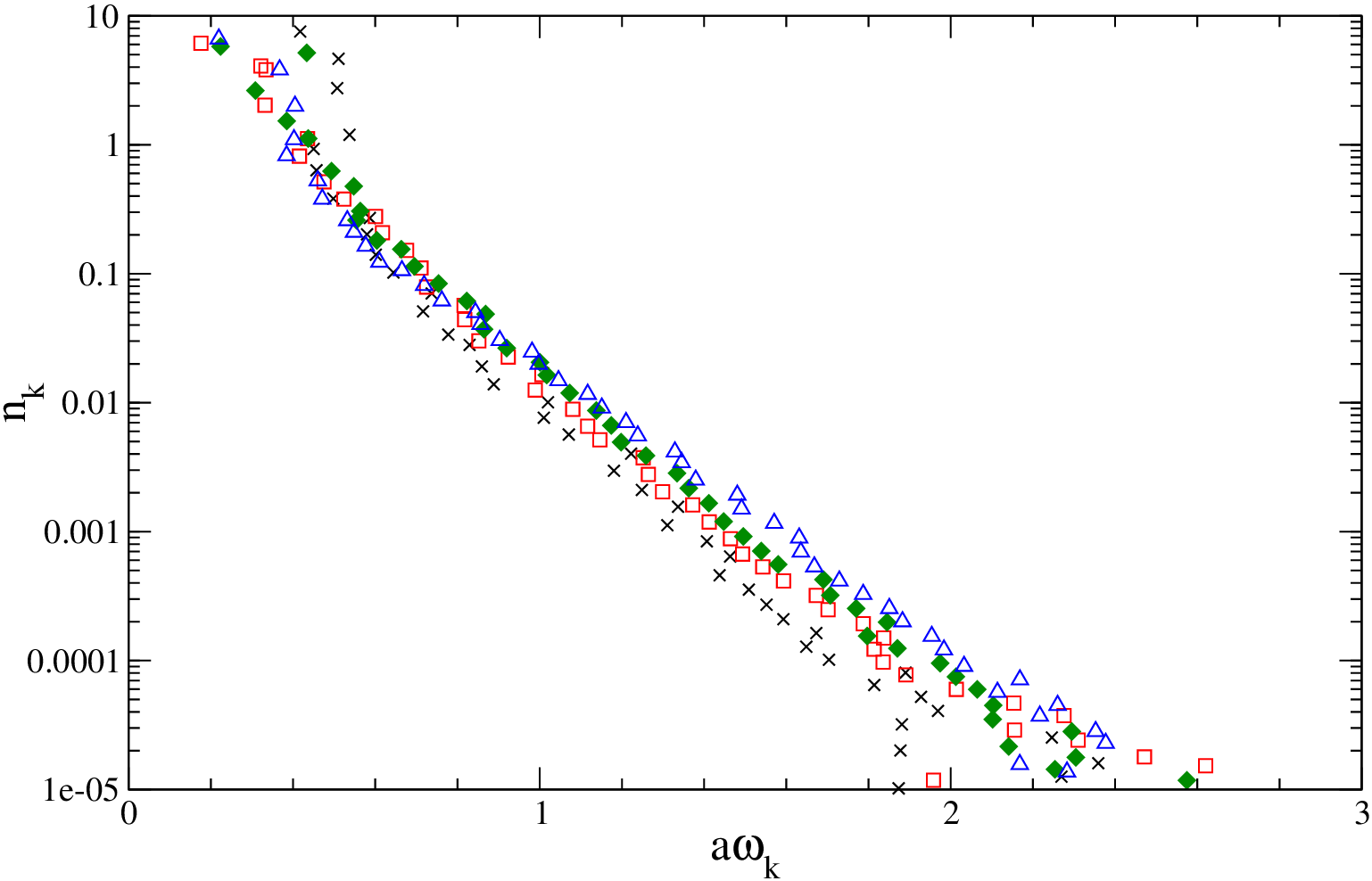}
}
\includegraphics*[width=0.49\textwidth]{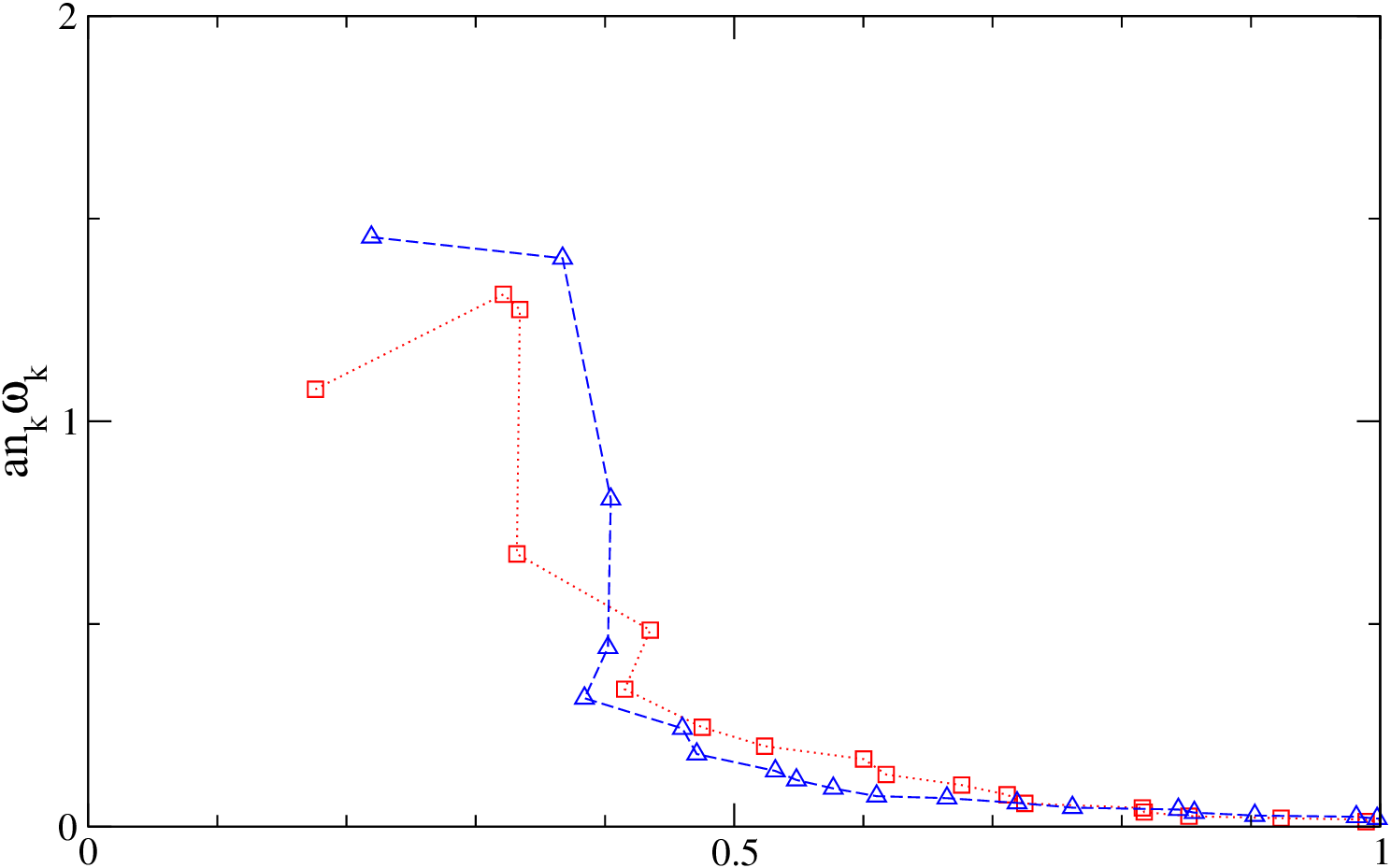}
\includegraphics*[width=0.49\textwidth]{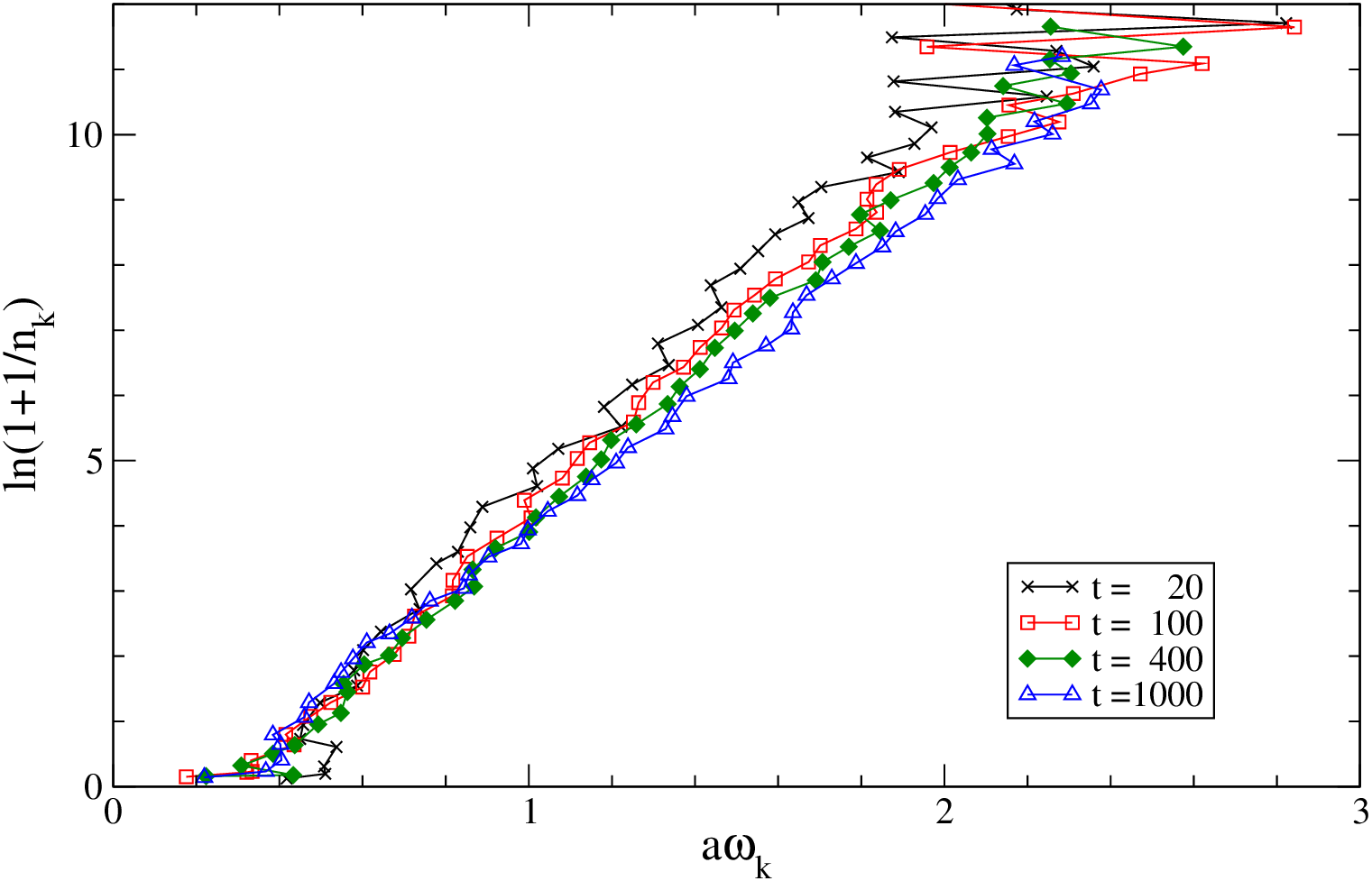}
\caption{
$n_k$ (top), $n_k\om_k a$ (bottom left) and $\ln (1+1/n_k)$ (bottom
right) versus $\om_k a$, at times $t m_H = 20$, 100, 400, 1000.
}
\end{figure}
At later times $20 \leq t m_H \leq 1000$ the distribution $n_k$ does not change
much, as can be seen in Figure~4. At $k=0$, the particle number 
stays roughly constant, $6 \lesssim n_0 \lesssim 7$, while the tail of the
distribution appears to rise slowly. There is little power beyond
$a\om_k = am_H = 0.35$  (top plot).

The data do not yet show clear signs of classical equilibration,
$n_k\to T/\om_k$.
One expects a plateau to develop in $n_k\om_k$, but the
`plateau' in the lower-left plot does not look very convincing yet
as it involves only two $k$-values.
\begin{figure}
\center{
\includegraphics*[width=0.70\textwidth]{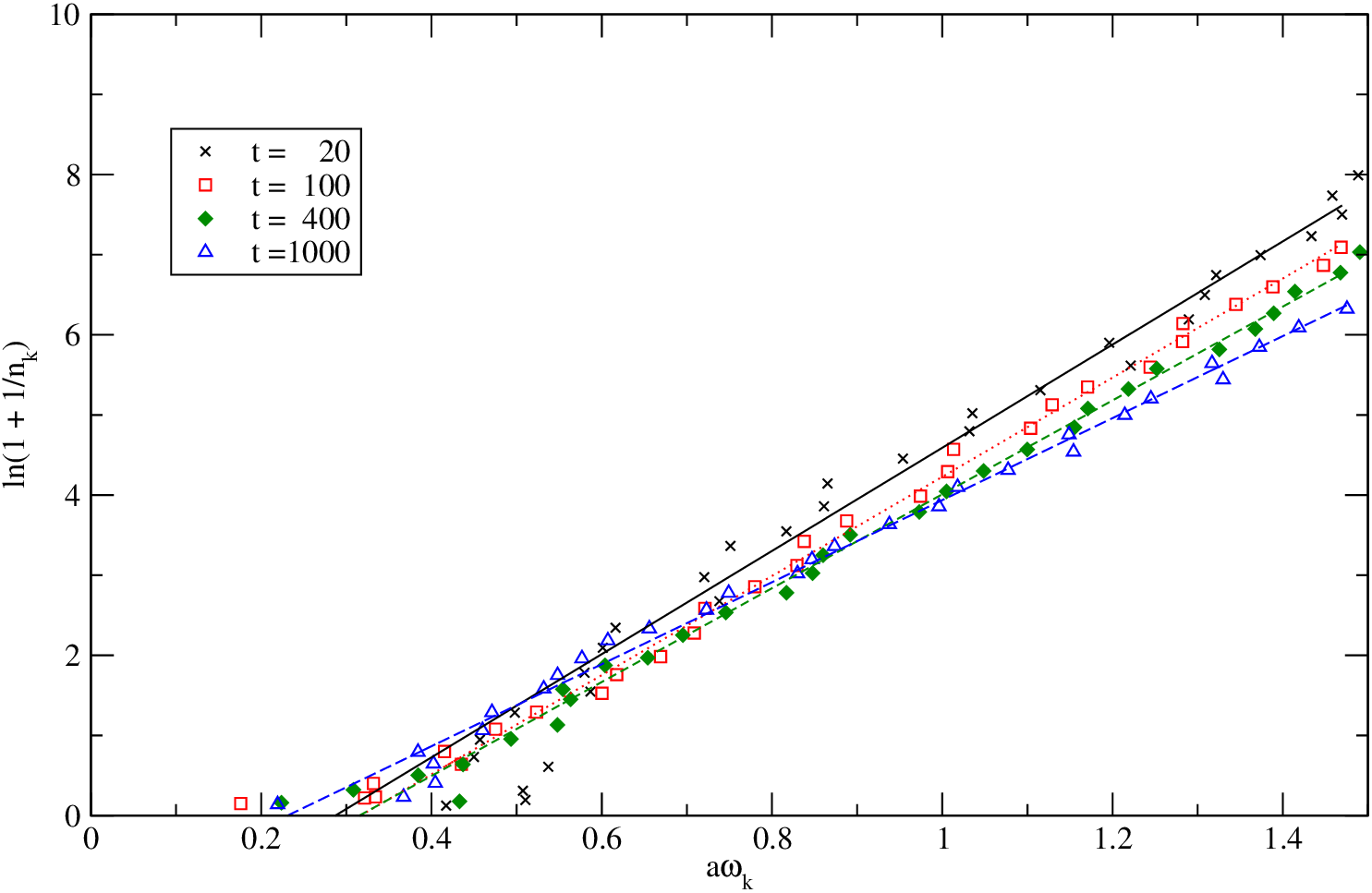}
}
\caption{
Linear fits $\ln (1+1/n_k) \approx (\om_k-\mu)/T$ resulting in an effective
temperature $T/m_H = 0.44$, 0.46, 0.49, 0.56, and effective chemical potential
$\mu/m_H = 0.8$, 0.9, 0.9, 0.7, respectively for $tm_H = 20$,
100, 400 and 1000.
}
\end{figure}
The lower-right plot in Figure 4 shows
$\ln(1+n_k^{-1})$,
which for a Bose-Einstein distribution would have the linear form
$(\om_k-\mu)/T$.
The data are roughly compatible with this behavior, the fits in Figure 5
indicate an effective temperature that slowly increases as the higher
momentum modes get more occupied,
$T/m_H = 0.44 \to 0.56$, with a hint of a decreasing effective chemical
potential $\mu\approx 0.8 \to 0.7$,
as $t$ increases from 20 to 1000 $m_H^{-1}$.

\section{Conclusion}
After $t\approx 20\, m_H^{-1}$,
the
particles produced by the instability settle
into an effective temperature of about $0.5\, m_H$.
Subsequent evolution is slow in the SU(2)-Higgs model,
in the classical approximations used here:
the modes are massive and the effective interactions appear to be
weak at the BE-temperature $T\approx 0.5\, m_H$.
The Higgs particles have roughly the same effective temperature.

The initial conditions for this simulation do not put power into the
high-momentum modes near the cutoff \cite{ST,Smit:2001qa}.
There is furthermore little draining of power from the low- to high-momentum
modes, so Rayleigh-Jeans effects are negligible.
Lattice artefacts are also under control, since the distribution function
drops exponentially fast and there is little power in the gauge-field
modes beyond $ak = 0.5$.
In particular, it makes sense to use a simple lattice version \cite{ST} of
an effective CP-violating term 
$\kp\Tr\ph^{\dagger}\ph F_{\mu\nu} \tilde F^{\mu\nu}$
in the lagrangian.\\ 

\noindent {\bf Acknowledgments}.\\
This work was supported in part by FOM.



\end{document}